\documentclass[journal]{IEEEtran}
\usepackage{stfloats}

\usepackage{enumerate}
\usepackage{algorithm,algpseudocode}
\usepackage{setspace,amsmath,latexsym,cite,amssymb,epsfig,amsfonts}
\usepackage{url,cite}
\usepackage{graphicx}
\usepackage{psfrag}
\usepackage{footmisc}
\usepackage{multirow}
\usepackage{color}
\usepackage{multicol}
\usepackage{mathtools}
\usepackage{subcaption}
\usepackage{graphicx}
\usepackage{amssymb}
\usepackage{amsmath}
\usepackage{epstopdf}
\usepackage{geometry}
\usepackage{float}
\usepackage{balance}
\usepackage{makecell}
\usepackage{changepage}

\algnewcommand\algorithmicforeach{\textbf{for}}
\algdef{S}[FOR]{For}[1]{\algorithmicforeach\ #1\ \algorithmicdo}

\twocolumn

\makeatother

        \makeatletter
        \def\fps@eqnfloat{!t}
        \def\ftype@eqnfloat{4}
        
        \newenvironment{eqnfloat*}
               {\@dblfloat{eqnfloat}}
               {\end@dblfloat}
        \makeatother
\allowdisplaybreaks[4]

\title{\huge Deep Reinforcement Learning-Based Resource Allocation\\for Hybrid Bit and Generative Semantic Communications\\in Space-Air-Ground Integrated Networks}
\author{Chong Huang, \IEEEmembership{Member, IEEE}, Xuyang Chen, \IEEEmembership{Student Member, IEEE}, Gaojie Chen, \IEEEmembership{Senior Member, IEEE}, \\Pei Xiao, \IEEEmembership{Senior Member, IEEE}, Geoffrey Ye Li, \IEEEmembership{Fellow, IEEE}, and Wei Huang
\thanks{C. Huang and P. Xiao are with 5GIC \& 6GIC, Institute for Communication Systems (ICS), University of Surrey, Guildford, GU2 7XH, United Kingdom, Email: \{chong.huang, p.xiao\}@surrey.ac.uk.}
\thanks{X. Chen is with the College of Electronics and Information Engineering, Shenzhen University, Shenzhen 518060, China, and also with the 5GIC \& 6GIC, Institute for Communication Systems, University of Surrey, GU2 7XH Guildford, U.K, Email: chenxuyang2021@email.szu.edu.cn.}
\thanks{G. Chen and W. Huang are with School of Flexible Electronics (SoFE) \& State Key Laboratory of Optoelectronic Materials and Technologies, Sun Yat-sen University, Guangdong, 510220, China. Email: gaojie.chen@ieee.org, huangw323@mail.sysu.edu.cn. (Corresponding author: G. Chen)}
\thanks{G. Y. Li is with the Department of Electrical and Electronic Engineering, Imperial College London, London SW7 2BX, U.K. E-mail: geoffrey.li@imperial.ac.uk.}
}

\begin{document}
\captionsetup[figure]{name={Fig.},labelsep=period}

\begin{singlespace}
\maketitle

\end{singlespace}

\thispagestyle{empty}
\begin{abstract}
In this paper, we introduce a novel framework consisting of hybrid bit-level and generative semantic communications for efficient downlink image transmission within space-air-ground integrated networks (SAGINs). The proposed model comprises multiple low Earth orbit (LEO) satellites, unmanned aerial vehicles (UAVs), and ground users. Considering the limitations in signal coverage and receiver antennas that make the direct communication between satellites and ground users unfeasible in many scenarios, thus UAVs serve as relays and forward images from satellites to the ground users. Our hybrid communication framework effectively combines bit-level transmission with several semantic-level image generation modes, optimizing bandwidth usage to meet stringent satellite link budget constraints and ensure communication reliability and low latency under low signal-to-noise ratio (SNR) conditions. To reduce the transmission delay while ensuring reconstruction quality for the ground user, we propose a novel metric to measure delay and reconstruction quality in the proposed system, and employ a deep reinforcement learning (DRL)-based strategy to optimize resource allocation in the proposed network. Simulation results demonstrate the superiority of the proposed framework in terms of communication resource conservation, reduced latency, and maintaining high image quality, significantly outperforming traditional solutions. Therefore, the proposed framework can ensure that real-time image transmission requirements in SAGINs, even under dynamic network conditions and user demand.
\end{abstract}


\section{Introduction}
In this section, we introduce the background of semantic communications and space-air-ground integrated network (SAGIN), along with related works. We then summarize the main contributions of this work.

\subsection{Background}
As indicated in \cite{9508471}, satellite communications have become a key component in the future sixth-generation (6G) network. Satellites can provide comprehensive access coverage and reliable communication services for ground users, especially in remote areas and oceans where traditional ground infrastructure deployment is difficult or costly. Currently, satellite projects, such as Starlink \cite{10098597} and OneWeb \cite{10551683} are developing rapidly to provide communication services for remote areas and emergency situations. Moreover, SAGINs have attracted significant attention in the current wireless communication research \cite{8368236}. The air layer can extend and relay communications between satellites and ground users, so that the network can be rapidly configured to quickly respond to dynamic changes in environmental conditions or network needs, such as natural disasters and emergency communications. In recent years, SAGINs have played a key role in many scenarios such as disaster monitoring, remote sensing, and real-time surveillance.

However, the link budget problem remains a major challenge in satellite communications \cite{9295418}. For services such as remote sensing data transmission, real-time ground monitoring, emergency data transmission, and big data forwarding, the transmission capacity of satellite communications is constantly challenged. Based on \cite{MONTIERI2021108529}, visual information, such as images and videos, occupies more than 66\% of the capacity of wireless communications, which can cause significant delay in satellite communications, especially when the signal-to-noise ratio (SNR) is at a low level. Therefore, it is urgent to develop technologies that can reduce the transmission resources required for visual data.

To achieve this goal, semantic communication has become a key technique for 6G communications due to its high spectrum efficiency \cite{10183794}. Unlike traditional communications that consider bit transmission rate, semantic communication exploits semantic decoders and encoders to handle semantic meaning or information features directly, which can greatly reduce the required spectrum resources. Therefore, semantic communication is very suitable for solving the problem of limited satellite link budget.

Although semantic communications can greatly reduce the consumption of spectrum resources, extracting information features at the transmitter and reconstructing them at the receiver may cause information loss \cite{9530497}. Some transmission tasks are highly sensitive to delay while others are sensitive to information quality \cite{chen2025communication}. Therefore, balancing the quality of reconstructed information, spectrum resource consumption, and the impact of transmission and computational delays will be an important research direction in semantic communication.

\subsection{Related Work}\label{sec:RW}
Satellites are good at long-distance and wide-area coverage communications, particularly in remote regions and oceans where terrestrial networks are inaccessible. There have been lots of research efforts in this area \cite{9110855,9628071,10463093,10516308}. In \cite{9110855}, massive multiple-input multiple-output (MIMO) has been utilized to enhance the transmission rate in low earth orbit (LEO) satellite communication networks. To improve the transmission rate and reduce the computational complexity, a precoding vector based on deep learning has been designed in downlink MIMO LEO networks \cite{9628071}. In \cite{10463093}, stochastic geometry has been utilized to model the distance distributions and interference characterizations, and improve the satellite coverage probability. The deep reinforcement learning (DRL)-based beam management scheme in \cite{10516308} was proposed to enhance the achieved data rate for LEO communications. To further explore the advantages of aerial platforms in satellite-terrestrial communications, many research works have considered the three-tiered space-air-ground communication architecture in future 6G \cite{9177315,9701875,10440193,10704927,HFLSAGIN24}. In \cite{9177315}, the closed-form for the outage performance in the SAGIN scenarios has been derived. To maximize the social welfare, the queuing game model has been adopted to design the dynamic handover transmission control scheme in the SAGIN system \cite{9701875}. To reduce the energy cost and task delay of multi-access edge computing (MEC), a decision-assisted DRL algorithm has been proposed in a SAGIN with cloud servers \cite{10440193}. The traffic-aware offloading framework in \cite{10704927} can predict data traffic and optimize resource allocation in SAGINs. To enhance the training performance in hierarchical federated learning frameworks, aerial platforms and LEO satellites in SAGINs have been utilized as edge servers and cloud servers in \cite{HFLSAGIN24}.

On the other hand, many research works have incorporated semantics into the scope of future wireless communication technologies \cite{9763856,9450827,9953076,10508293}. In \cite{9763856}, the channel resource and the size of semantic information have been optimized to enhance the semantic spectral efficiency for text transmissions. The deep learning-based semantic communication framework in \cite{9450827} can enhance the signals metrics in multimedia transmissions. In \cite{9953076}, deep learning has been adopted to extract and reconstruct the original information for image semantic communications. To improve the quality of experience (QoE) of users, the semantic systems in \cite{10508293} help the MEC system to reduce the offloading pressure. Considering the limitations of satellite link budget, semantic communications have gained significant attention in the current satellite communication research \cite{10445211,10436114}. In \cite{10445211}, the adaptive framework and the computational task processing scheduling scheme can optimize the resource in semantic satellite communications and decrease the delay and energy cost. To reduce the age of information (AoI) in deep space transmissions, the semantic correlations is exploited to enhance error tolerant capability of semantic communications \cite{10436114}.

However, in practical wireless environments, satellite communications often require more flexible transmission schemes to accommodate tasks with varying demands for delay and information quality. The hybrid bit and semantic text communication framework in \cite{10477313} can mitigate the disadvantages of each other. However, a text transmission framework is insufficient for future 6G since images and videos account for the majority of data traffic in wireless communications. In \cite{yu2024hybrid}, a hybrid bit and semantic transmission system was investigated for image transmissions. Nevertheless, considering the dynamic wireless communication environments and task requirements, the hybrid transmission of solely text or images with bits is insufficient to meet the demands of future satellite communication systems. Moreover, with the development of generative artificial intelligence (GAI), offering more diverse semantic transmission options, such as image-text embedding and varying sizes of image semantic information for image transmissions, would be beneficial in future 6G networks. Furthermore, for complex and high-resolution images, more powerful algorithms and tools are needed for semantic extraction and reconstruction.

\subsection{Motivation and Contributions}
In this paper, we explore the hybrid bit and semantic communication framework in SAGINs. We propose a hybrid communication framework, where the semantic component includes text + image and multiple types of image semantic transmission. This framework is designed to meet various transmission and delay requirements while adhering to the satellite link budget constraints. Moreover, we investigate the impact of resource allocation on downlink transmissions, including UAV dynamic planning, UAV-satellite and UAV-ground user pairing, and inter-satellite transmission relationships within the SAGIN. Our goal is to minimize transmission resource consumption while satisfying the transmission task's delay or reconstruction quality requirements under limited satellite link budgets. Considering the complexity of the proposed problem, we apply a DRL algorithm to solve this optimization challenge. The main contributions of this work are summarized as follows.

\begin{itemize}
  \item We propose a novel framework for hybrid semantic and bit communication in SAGINs. Considering the satellite coverage constraints, we introduce inter-satellite links (ISLs) to enable information forwarding between LEO satellites when some satellites are outside the coverage area for UAVs and ground users. In the air layer, UAVs serve as relay points between satellites and ground users for satellite downlink communications. To mitigate the effects of channel fading in air-ground communications, we introduce UAV trajectory planning and user pairing strategies.

  \item Our framework not only includes bit transmission to ensure high-quality information transmission but also offers multiple semantic image transmission modes, such as text-to-image mode and text \& image embeddings mode, to balance the performance and task completion time. Moreover, we introduce a new communication metric, semantic communication efficiency (SCE), which measures task delay and information quality in semantic communications.

  \item The optimization variables include UAV-ground user pairing, UAV-satellite pairing, ISL selection, UAV dynamic planning, and semantic transmission mode selection. To solve this complicated optimization problem, we employ a distributed soft actor-critic (DSAC) algorithm to maximize long-term benefits in SAGIN.

  \item Simulation results present the superiority of our proposed hybrid bit and semantic communication framework and DRL-based optimization algorithm. Moreover, we analyze the effects of the SCE metric under different weights to show its adaptability to various requirements.
\end{itemize}

The rest of this paper is summarized as follows: In Section \ref{sec:sm}, the SAGIN system model, including the communication model and satellite coverage model are introduced. The hybrid bit and semantic communication framework, the metric SCE, and problem formulation are described in \ref{sec:semantic}. In Section \ref{sec:DSAC}, the DRL-based resource allocation algorithm is proposed. Section \ref{sec:sim} shows the advantages of the proposed framework and algorithm in the simulation. Finally, Section \ref{sec:con} concludes this paper.

\section{SAGIN System Model} \label{sec:sm}
\begin{figure}[t!]
  \centering
  \centerline{\includegraphics[width=0.47\textwidth]{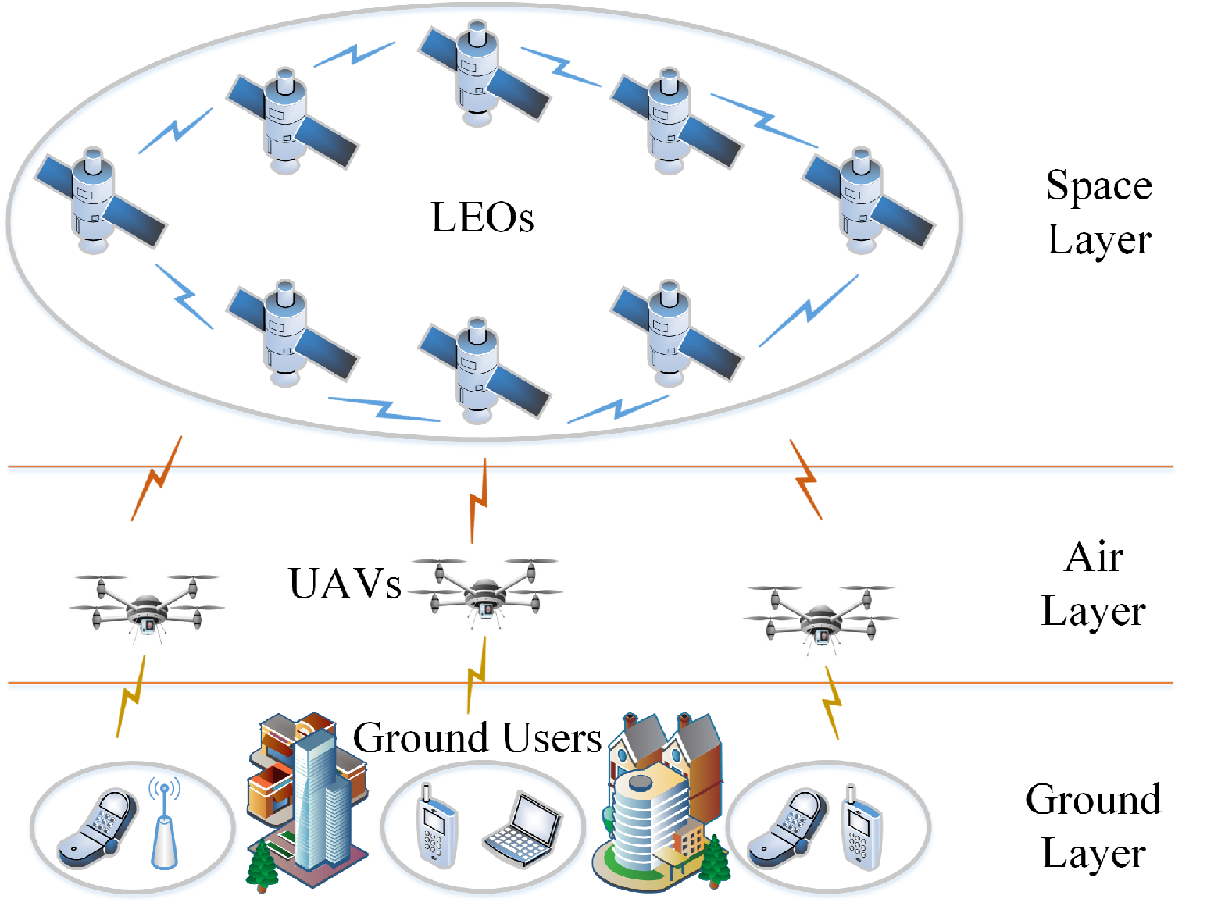}}
 \caption{System model of SAGINs.} \label{fig:SM}
\end{figure}

In this paper, the SAGIN framework is considered to provide global communication coverage for ground users. Due to the coverage limitation of satellites and hardware limitations of ground users, UAVs will provide relay service in SAGINs. As shown in Fig. \ref{fig:SM}, there are $K$ LEO satellites $S_k$ ($k \in \mathcal K = \{1, 2, ..., K\}$), $N$ UAVs $U_n$ ($n \in \mathcal N = \{1, 2, ..., N\}$), and $M$ ground users $F_m$ ($m \in \mathcal M = \{1, 2, ..., M\}$). The 3D coordinate of UAV $U_n$ is denoted as $o_{n}(t) = \{x_{n}(t), y_{n}(t), z_{n}(t)\}$ ($o_n \in \mathcal O = \{o_1, o_2, ..., o_N\}$) at time slot $t$.

\begin{itemize}
  \item \textbf{Space Layer}: The space layer includes $K$ LEO satellites, as image data sources in SAGINs. The LEO satellites can collect high-resolution images of the Earth's surface, or receive image data from other LEO and ground users. Considering the long distance between space and air layers, efficient pre-processing and compression techniques are needed to reduce data volume and latency and mitigate link budget limitations prior to transmitting the data to the air layer.

  \item \textbf{Air Layer}: The air layer includes $N$ UAVs acting as relays between LEO satellites and ground users. Considering that ground users may be widely dispersed, UAVs need to optimize their trajectories to search the optimal coverage area for ground users and ensure efficient resource allocation. Thus, the air layer plays a key role in connecting LEO satellites with terrestrial users, mitigating signal attenuation issues, and providing reliable communication service. However, the LEO-UAV communication is subject to time constraints due to the limitation of minimum elevation angle in satellite communications.

  \item \textbf{Ground Layer}: In this layer, there are $M$ ground users, which may include base stations or mobile devices. The ground users play as receivers in communications of SAGINs, since they operate in emergency environments where traditional communication infrastructure is not deployed. The ground users need to decode the received data from air layer and then reconstruct the images. These images can be used for decision-making processes in applications, such as disaster response, precision agriculture, and urban planning.
\end{itemize}

Thus, the LEO satellites needs to send images, such as remote sensing data, to ground users based on transmission task requirements. Since satellites cannot connect directly with ground users in SAGINs, it first sends the data to a UAV, which then forwards it to the respective ground user. Considering that ground users may be widely dispersed, the trajectory planning for UAVs is important to ensure optimal service for ground users. Besides, the LEO-UAV communication is subject to time constraints due to the minimum elevation angle required for reliable connectivity.

\subsection{Communication Model}
For satellite-UAV downlink communications, we assume that each UAV can establish a connection with a single LEO satellite during any time slot. Binary indicator $r_{k,n} = 1$ represents that satellite $S_k$ communicates with UAV $U_n$, otherwise $r_{k,n} = 0$. The channel coefficient between satellite $S_k$ and UAV $U_n$ is given by

\begin{equation}\small\label{eq4}
\hat{h}_{k,n} = \frac{\sqrt{\varsigma_k}\lambda}{4 \pi d_{k,n}} e^{j {\nu_k}},
\end{equation}
where $\varsigma_k$ presents the antenna gain of LEO $S_k$, $\lambda$ indicates the wavelength, $d_{k,n}$ denotes the distance between $S_k$ and $U_n$, and $\nu_k$ represents the phase component of $S_k$. In the proposed SAGIN, the impact of outdated channel state information (CSI) is introduced due to the significant distance between space and air layers. The outdated CSI is modeled as \cite{10287142}

\begin{equation}\small\label{eq5}
h_{k,n} = \delta \hat{h}_{k,n} + \sqrt{1 - \delta^2} \hat{g}_{k,n},
\end{equation}
where $\delta = \hat{\omega} (2 \pi {\hat{D}}_{k,n} T_{k,n})$, with $\hat{\omega}$ being the Bessel function of the first kind of order 0, ${\hat{D}}_{k,n}$ denoting the maximum Doppler frequency, and $T_{k,n}$ representing the transmission delay. Term $\hat{g}_{k,n}$ is a complex Gaussian random variable with variance $\hat{h}_{k,n}$. Consequently, the transmission rate from satellite $S_k$ to UAV $U_n$ can be expressed as

\begin{equation}\small\label{eq7}
C_{k,n} = B_{k,n} {\rm{log_{2}}} \left( 1 + \frac{P_{S_k} |h_{k,n}|^2}{{{\sigma}_{n}^2}} \right),
\end{equation}
where $B_{k,n}$ is the bandwidth for the downlink transmission between $S_k$ and $U_n$, $P_{S_k}$ is the transmit power of satellite $S_k$, and ${{\sigma}_{n}^2}$ denotes the additive white Gaussian noise (AWGN) at UAV $U_n$. Interference is not present during this stage because LEOs operate on different channels.

Moreover, ISLs are utilized for communication between satellites within this SAGIN framework. Due to the coverage limitation of LEO satellites, one satellite can send task data to other LEOs to let other LEOs help complete tasks if it is out of service due to the limitation of minimum elevation angle but still has transmission tasks. The data rate of the ISL between two satellites, $S_i$ and $S_j$, can be expressed as \cite{ISL_TWC21}

\begin{equation}\small\label{eq8}
C_{i,j} = B_{i,j} {\rm{log_{2}}} \left( 1 + \frac{P_{S_i} |\eta_{\rm max}|^2}{\zeta \chi B_{i,j} \left(\frac{{4 \pi d_{i,j} f_{\rm I}}}{v_c}\right)^2} \right),
\end{equation}
where $B_{i,j}$ represents the bandwidth of the channel from $S_i$ to $S_j$, $P_{S_i}$ is the transmit power of satellite $S_i$, and $d_{i,j}$ is the distance between the satellites. The parameter $\eta_{\rm max}$ is the peak antenna gain of $S_i$ in the transmission direction, $\zeta$ is the Boltzmann constant, $\chi$ is the thermal noise, $f_{\rm I}$ denotes the ISL carrier frequency, and $v_c$ is the speed of light. We assume $y_{i,j} = 1$ denotes LEO $S_i$ sends the data to $S_j$, otherwise $y_{i,j} = 0$. Interference is avoided by employing narrow beam antennas and precise beam steering as in \cite{ISL_TWC21}.

For air and ground layers, we assume that each ground user is supported by only one UAV, while a single UAV can serve multiple ground users. We assume a binary indicator $l_{n,m} = 1$ denotes that UAV $U_n$ and ground user $F_m$ are connected, otherwise $l_{n,m} = 0$. The channel coefficient $h_{n,m}$ between UAV $U_n$ and ground user $F_m$ can be expressed as

\begin{equation}\small\label{eq1}
h_{n,m}= \sqrt{\frac{\mu}{\mu+1}} \bar{H}_{n,m}+\sqrt{\frac{1}{\mu+1}} \hat{H}_{n,m},
\end{equation}
where $\mu$ denotes the Rician factor. The terms $\hat{H}_{n,m} = \hat{\mathbb{g}}_{n,m} d_{n,m}^{-{\kappa_N}/2}$ and $\bar{H}_{n,m} = \bar{\mathbb{g}}_{n,m} d_{n,m}^{-{\kappa_L} / 2}$ represent the non-line-of-sight (NLoS) and line-of-sight (LoS) channel coefficients, respectively. Here, $d_{n,m}$ is the distance between UAV $U_n$ and ground user $F_m$, while $\kappa_N$ and $\kappa_L$ denote the path loss exponents for the NLoS and LoS channels, respectively. In the NLoS scenario, $\hat{\mathbb{g}}_{n,m}$ represents a zero-mean, unit-variance Gaussian fading component.

Thus, the transmission rate between UAV $U_n$ and ground user $F_m$ can be formulated as

\begin{equation}\small\label{eq2}
C_{n,m} = B_{n,m} {\rm{log_{2}}} \left( 1 + \frac{P_{U_n} |h_{n,m}|^2} { \sum_{i=1,i \neq n}^{N} P_{U_i} |h_{i,m}|^2 + {{\sigma}_{n}^2} } \right),
\end{equation}
where $B_{n,m}$ indicates the bandwidth of the channel from $U_n$ to $F_m$, $P_{U_n}$ is the transmit power of UAV $U_nk$.

\subsection{Satellite Coverage Model}
\begin{figure}[t!]
\centering
\centerline{\includegraphics[width=0.33\textwidth]{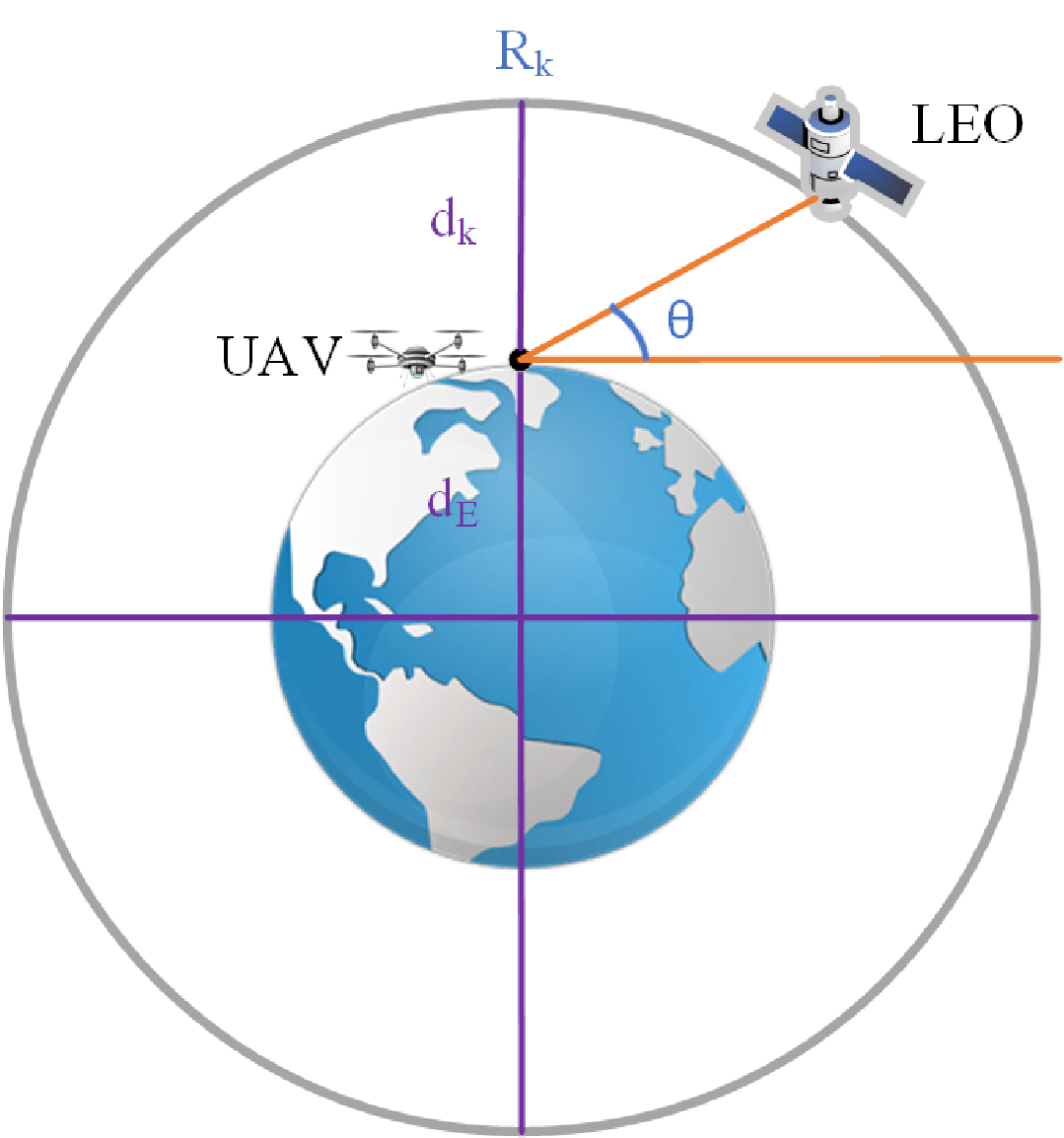}}
\caption{Coverage area of LEOs for UAV receivers.} \label{fig:coverage}
\end{figure}

Since LEO satellites maintain high-speed movement around the Earth, the access time each LEO satellite can provide to a specific ground receiver, such as a UAV, is limited over extended periods. To describe the impact of LEO satellite coverage range on the downlink communication link budget, we consider the effect of satellite access time on air platforms. As shown in the Fig. \ref{fig:coverage}, the path length traversed by satellite $S_k$ while providing access service to UAV $U_n$ is \cite{9344666}

\begin{equation}\small\label{eqc1}
R_k = 2 (d_E + d_k) \bigg(\arccos \Big(\frac{d_E}{d_E+d_k} \cos \tau \Big) - \tau \bigg),
\end{equation}
where $d_E$ represents the radius of the Earth, $d_k$ is the altitude of the LEO satellite $S_k$, and $\tau$ is the minimum elevation angle of LEO satellites. Based on this, the total duration for communication between the LEO satellite $S_k$ and UAV $U_n$ can be expressed as

\begin{equation}\small\label{eqc2}
T_k = \frac{R_k}{v_k},
\end{equation}
where $v_k$ is the orbital velocity of $S_k$. In SAGIN systems, due to the varying orbital positions of LEO satellites, each LEO has a different remaining access time for the same UAV at any given time slot. Furthermore, since the Earth’s radius and satellite altitude are numerically far greater than the UAV’s altitude, the UAV altitude can be neglected when calculating the LEO’s remaining access time as in \eqref{eqc2}.

\section{Hybrid bit \& Semantic Communication Framework and Problem Formulation} \label{sec:semantic}
In this section, we present the proposed hybrid bit and semantic communication framework and its training methodology. Moreover, we define a new metric to indicate the tradeoff between delay and semantic quality, upon which we base our optimization formulation.

\begin{figure*}[t!]
    \centering
    \includegraphics[width=0.9\textwidth]{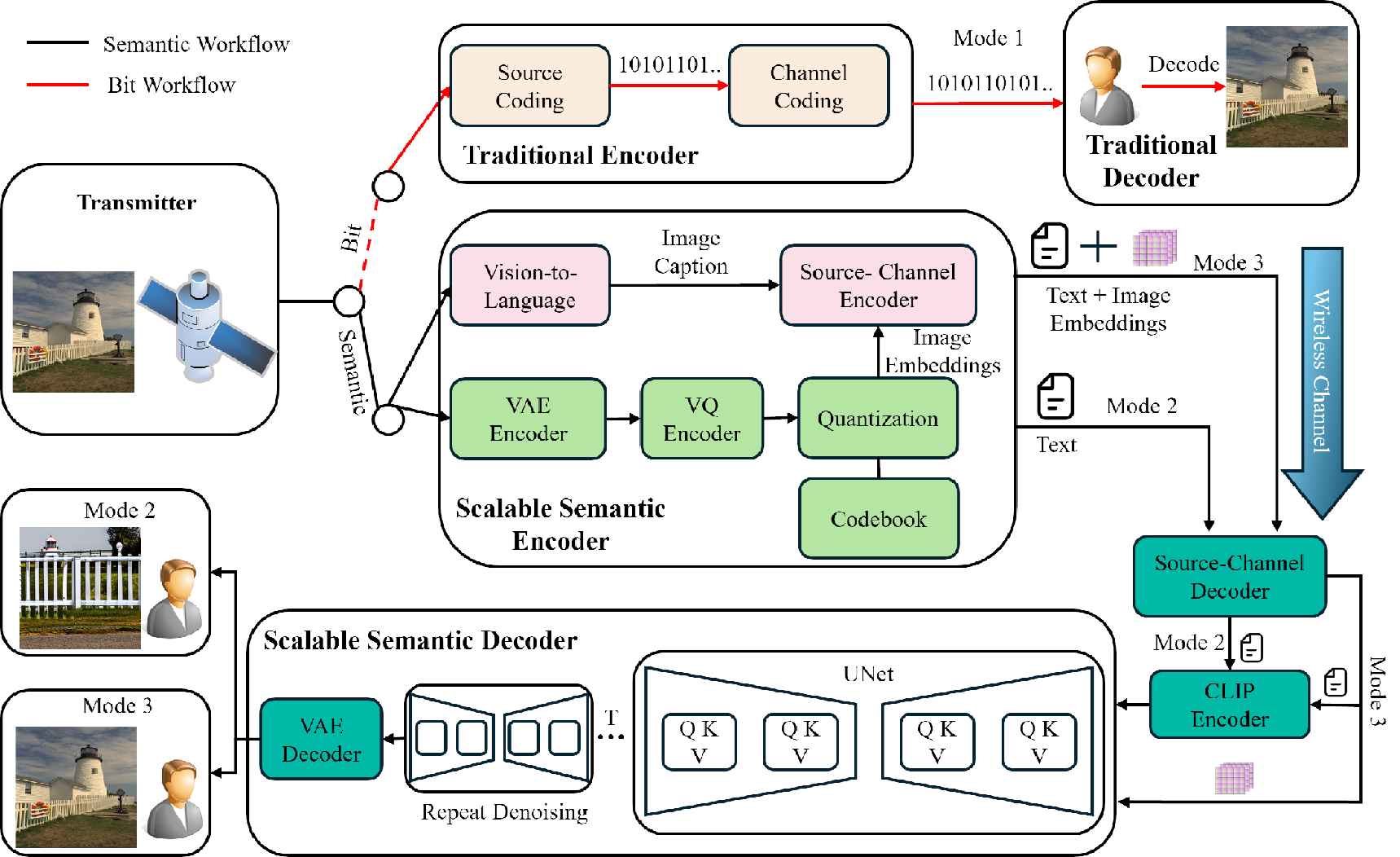}
    \caption{The Proposed Scalable Semantic Transmission Framework.}
    \label{sys-framework}
\end{figure*}

\subsection{Semantic Communication Framework}
Semantic communications focus on transmitting the features of messages rather than sending every bit precisely to the receiver, which can still work well in the low SNR environments. In this work, we assume the LEO satellites is equipped with semantic encoders and ground users have semantic decoders. To accommodate users with varying computational capabilities and communication demands, the proposed scalable semantic transmission model supports the following three image transmission methods:
\begin{enumerate}
    \item \textbf{Mode 1}: Traditional image source-channel coding methods, such as joint photographic experts group (JPEG) combined with low-density parity-check (LDPC), which require the least computational resources, offer the best restoration quality but have high transmission volumes and poor scalability.
    \item \textbf{Mode 2}: Extracting and transmitting image text to directly generate the original image. This method requires substantial computation and offers the best compression performance, but the restoration quality is subpar.
    \item \textbf{Mode 3}: Extracting both image text and image features and using this information to reconstruct the image. This method is computationally intensive, provides a moderate compression ratio, and has certain scalability by adjusting the size of the image features.
\end{enumerate}

As shown in Fig. \ref{sys-framework}, LEO satellites can either utilize traditional source-channel coding for image transmission or opt for scalable semantic transmission. The scalable semantic encoder on the edge includes a vision-to-language extraction module, a variational autoencoder (VAE) \cite{vae}, a vector quantization encoder (VQE), a codebook, and a source-channel encoder. The vision-to-language extraction module uses BLIP-2 \cite{blip2} to obtain a comprehensive description of the image. Text produced by BLIP-2 is encoded into textual features $\boldsymbol{w}$ by the contrastive language-image pre-training (CLIP) encoder \cite{clip}. The image $\boldsymbol{x}$ is encoded by the VAE into a Gaussian-like distribution, outputting a series of means and variances. These values are then fed into the VQE, and based on the codebook, the closest codeword $\boldsymbol{z}$ is selected, which is then transmitted after source-channel encoding. The output spatial resolution of the VQE supports multi-level adjustments, where lower spatial resolutions consume fewer transmission bits to accommodate different communication rate requirements. During each training iteration, the image and its caption are encoded by VAE and CLIP into $\boldsymbol{x_0}$ and $\boldsymbol{w}$. Afterward, a forward diffusion Markov process is performed. Specifically, a scheduler gradually adds Gaussian noise at each time step $t\in[0,T]$ until the latent descriptions $\boldsymbol{x_0}$ become pure Gaussian noise $\boldsymbol{x_T}$, i.e.,

\begin{equation}
    \boldsymbol{x}_t = \sqrt{\overline{\alpha}_t}\boldsymbol{x}_0 + \sqrt{1-\overline{\alpha}_t}\epsilon,
\end{equation}
where $\alpha_t=1-\beta_t$ and $\overline{\alpha}_t=\prod \limits_{i=1}^t\alpha_i$. $\{\beta_t\}^T_{t=1}$ is pre-defined by the scheduler, which controls the intensity of noise addition. $\epsilon\sim \mathcal{N}(0,\boldsymbol{I})$ is the added Gaussian noise with an identity matrix $\boldsymbol{I}$.

The scalable semantic decoder is deployed on the user side. Users restore clear images by performing a reverse diffusion process, also known as de-noising. The de-noising module utilizes a neural network, UNet \cite{sdm}, to predict the noise corresponding to each time step and restores the original image under the guidance of text and image embeddings. Thus, the de-noising process is represented as

\begin{equation}
\label{denoise}
    \boldsymbol{x}_{t-1} = \frac{1}{\sqrt{\alpha_t}}\left(\boldsymbol{x}_t-\frac{1-\alpha_t}{\sqrt{1-\overline{\alpha}_t}}\epsilon_{\boldsymbol{\theta}}\left(\boldsymbol{x}_t,t,\boldsymbol{z,\boldsymbol{w}}\right)\right),
\end{equation}
where $\epsilon_{\boldsymbol{\theta}}$ means the noise predicted by UNet, and $\boldsymbol{\theta}$ denotes the UNet parameters. At each layer of the UNet, image embeddings are concatenated with the noisy latent features $x_t$ of the current time step along the batch dimension, forming a hybrid feature tensor. The concatenated hybrid features interact with text prompt features via cross-attention, enabling the model to fuse global semantic constraints from text with local priors from image embeddings.

By iteratively applying Eq. (\ref{denoise}), the latent features of the image, $\boldsymbol{x_0}$, are generated. Ultimately, the high-definition source image is produced via VAE.

\begin{figure*}[t!]
    \centering
    \includegraphics[width=0.85\linewidth]{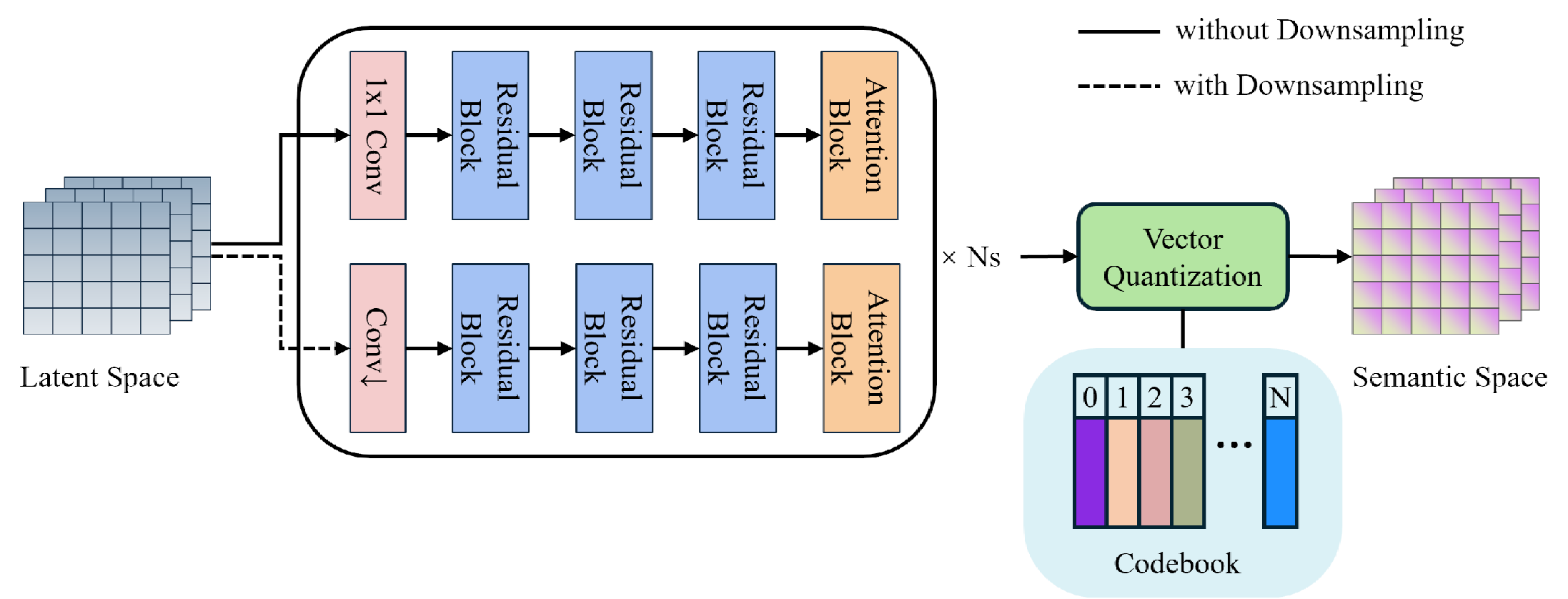}
    \caption{The VQE Architecture.}
    \label{VQE}
\end{figure*}

As illustrated in Fig. \ref{sys-framework}, semantic transmission offers two modes: transmitting the image description (Mode 2) or transmitting both the image description and its features (Mode 3), with the latter requiring more bandwidth but providing better recovery outcomes. Mode 2 adjusts the transmission rate according to the target rate by introducing VQE. Taking an input image of $512\times512$ as an example, the VAE first maps the image to a feature map of size $4\times64\times64$. Then the feature map is fed to VQE, which comprises a series of convolutional layers, and projects the feature map to a lower dimension, such as a spatial resolution of $h\times w$, which correlates with the desired target bitrate. The detailed architecture of the VQE is shown in Fig. \ref{VQE}. In the latent space, VQE further extracts semantic features through a series of convolutional blocks, residual blocks, and attention blocks. A convolution block with a stride of 2 is employed to reduce the spatial resolution when the target dimension is smaller than the current dimension; otherwise, a 1x1 convolution block is used. The attention block incorporates a gating mechanism that generates a weight distribution for the feature maps, selectively amplifying or suppressing input features by learning the gating signals. After passing through $N_s$ cascaded feature extractors, vector quantization is performed to obtain a more compact encoded representation, i.e., the semantic representations. The number of quantized bits is related to the pre-defined size of the codebook. When the spatial resolution is $h\times w$ and the codebook size is $V$, the output of VQE requires $hw\log_2 V$ bits for representation.

In the proposed SAGIN framework, a semantic encoder is deployed on each LEO satellite while a semantic decoder is deployed at each ground user. The UAVs in the aerial layer are responsible for relaying the semantic information extracted by the semantic encoder.

\subsection{Training Algorithm For Semantic Models}
We focus on the scalability of the model and the quality of image restoration. Therefore, VAE, BLIP-2, and CLIP use the most advanced pre-trained models and do not participate in model training. The training strategy for VQE is the same as that for VQ-VAE \cite{vq-vae}, aimed at finding the closest codeword in the codebook to approximate each vector of the current feature. For a feature $\boldsymbol{H_s}$ with spatial resolution $h\times w$ the quantization loss for each vector $\boldsymbol{h_s}$ mapped to the codebook is

\begin{equation}
\mathcal{L}_{\mathrm{VQ}}=\mathbb{E}_{\boldsymbol{h}_s}\left[\left\|s g\left(\boldsymbol{h_s}\right)-\boldsymbol{z_q}\right\|_2^2+\left\|s g\left(\boldsymbol{z_q}\right)-\boldsymbol{h_s}\right\|_2^2\right],
\end{equation}
where $sg(\cdot)$ is the stop-gradient operation, and $\boldsymbol{z_q}$ is the codeword of the codebook. Eq. (3) guides the image feature vectors $\boldsymbol{h_s}$ to converge towards codewords $\boldsymbol{z_q}$ in the codebook. When the transmitter and receiver share the same codebook, it is only necessary to transmit the codeword indices corresponding to all feature vectors. This allows for reconstruction at the receiver based on the codebook, greatly saving the bandwidth required to transmit image features. UNet is trained to predict the noise at the current timestep. For image restoration, each de-noising step requires conditional guidance. Specifically, the image embeddings output by VQE are concatenated with latent representations and fed into UNet while text description features guide the image generation in the desired direction through cross-attention layers. The loss function for UNet is characterized by mean-squared error (MSE):

\begin{equation} \mathcal{L}_{\mathrm{UNet}}=\mathbb{E}_{P_{\boldsymbol{x}}}\mathbb{E}_{P_{\boldsymbol{z},\boldsymbol{x_t}|\boldsymbol{x}}}\underset{\epsilon\sim \mathcal{N}(0,\boldsymbol{I})}{\mathbb{E}}\left|\left|\epsilon-\epsilon_{\boldsymbol{\theta}}(\boldsymbol{x_t},t,\boldsymbol{z},\boldsymbol{w})\right|\right|^2_2.
\end{equation}
Classifier-free guidance \cite{ho2021classifierfree}, which does not require training additional classifiers, is widely used in the generative methods based on diffusion models. During the inference phase, the noise prediction is defined as a weighted sum of two scenarios: with and without text conditions. It can be expressed as
\begin{equation}
\begin{aligned}
\hat{\epsilon}_{\boldsymbol{\theta}}&=\epsilon_{\boldsymbol{\theta}}\left(\boldsymbol{x}_t,\left(\boldsymbol{z}, \emptyset\right), t\right)+\lambda_s\left(\epsilon_{\boldsymbol{\theta}}\left(\boldsymbol{x}_t,\left(\boldsymbol{z}, \boldsymbol{w}\right), t\right)\right.\\&-\left.\epsilon_{\boldsymbol{\theta}}\left(\boldsymbol{x}_t,\left(\boldsymbol{z}, \emptyset\right), t\right)\right)
\end{aligned}
\end{equation}
where $\lambda_s$ is the guidance scale. The larger $\lambda_s$ is, the more closely the generated content matches the text description. To make classifier-free guidance feasible, we randomly omit the text conditions with a 10\% probability during each training iteration. The omitted text conditions are replaced with learnable text embeddings.

\subsection{Semantic Communication Efficiency Metric}
To evaluate the performance of the hybrid bit and semantic communication system within SAGINs, particularly under constrained link budgets, it is essential to balance delay and reconstruction quality. Traditional semantic communication metrics, such as standalone delay or image quality indicators, are insufficient to fully describe the trade-offs between reconstruction quality and transmission delay in resource-limited environments, thus introducing the need for a new evaluation metric. Furthermore, to meet efficiency and delay requirements for transmission tasks, the SAGIN must effectively transmit large volumes of image data within limited link budgets. Although the rapid development of GAI has enabled semantic communication systems to efficiently extract image features and substantially reduce the size of transmitted semantic information, current AI models for image and text generation require significant computational resources for complex semantic analysis and generation, thereby introducing notable computational delay in wireless communications. Consequently, we propose a new metric SCE based on delay and image reconstruction quality to comprehensively evaluate the delay-performance trade-off for hybrid bit and semantic-based communication systems in SAGINs. We assume there are $G$ tasks $\phi_g$ ($g \in {\mathcal G} =\{1, 2, ..., G\}$) in total, the SCE for a image transmission task $\phi_g$ can be expressed as
\begin{equation}\small\label{metric}
\varpi_g = \xi_{q} (Q_g - Q_{g,\rm min}) + \xi_{d} (D_{g,\rm max} - D_g),
\end{equation}
where $\xi_{q} \in (0,1)$ and $\xi_{d} \in (0,1)$ denote the weights of reconstruction quality and delay for task $\phi_g$, respectively, $\xi_{q} + \xi_{d} = 1$. $Q_g$ indicates the reconstruction quality metric, e.g. peak signal-to-noise ratio (PSNR). $D_g$ presents the delay of task $\phi_g$ when completing transmission. $Q_{g,\rm min}$ denotes the minimum requirement of generative image quality at the receiver, and $D_{g,\rm max}$ presents the maximum delay constraint for task $\phi_g$. To facilitate the representation of $Q_g$ and $D_g$, we normalize each of them to a range between 0 and 1 according to their respective scales. This metric allows us to better analyze how to balance delay and task quality within a complex multi-layer communication systems. In our transmission system, bit transmission can preserve nearly complete information fidelity but requires large transmission resource, thus leading to delays for transmission tasks. In contrast, semantic transmissions reduce the demand on link budgets by sacrificing some information quality, resulting in reduced delay values. Furthermore, the flexibly defined modes of semantic transmissions offer additional options within SAGIN for exploring such trade-offs.

\subsection{Problem Formulation}
In this paper, we consider a scenario where a LEO satellite may receive multiple transmission tasks at a time slot, each task requires the delivery of a specified image to a corresponding ground user. Thus, our objective is to maximize the average SCE of all transmission tasks. The optimization variables include UAV-user pairing, UAV trajectory planning, UAV-satellite pairing, whether to transmit through ISLs to other satellites, and the selection of hybrid bit and semantic transmission modes. The proposed optimization formulation is as
\begin{align}
    \bold{\rm (P1)}: &\max_{\mathcal{R}(t), {\mathcal O}(t), \mathcal{L}(t), \mathcal{Y}(t), \varepsilon(t)} \sum_{g=1}^{G} \varpi_g,\label{SecrecyFunc}\\
    {\rm s.t.}&~ \sum_{k=1}^{K} r_{k, n} \leq 1, \forall n \in {\mathcal N} \tag{\ref{SecrecyFunc}{a}}, \label{SecrecyFuncSuba}\\
    &\sum_{n=1}^{N} r_{k, n} \leq N, \forall k \in {\mathcal K} \tag{\ref{SecrecyFunc}{b}}, \label{SecrecyFuncSubb}\\
    &\sum_{n=1}^{N} l_{n, m} \leq 1, \forall m \in {\mathcal M} \tag{\ref{SecrecyFunc}{c}}, \label{SecrecyFuncSubc}\\
    &\sum_{m=1}^{m} l_{n, m} \leq M, \forall n \in {\mathcal N} \tag{\ref{SecrecyFunc}{d}}, \label{SecrecyFuncSubd}\\
    &y_{k} \leq 1, \forall k \in {\mathcal K} \tag{\ref{SecrecyFunc}{e}}, \label{SecrecyFuncSube}\\
    &v_n \leq v_{\max}, \forall n \in {\mathcal N} \tag{\ref{SecrecyFunc}{f}}, \label{SecrecyFuncSubf}\\
    &z_n \geq z_{\min} ~\&~ z_n \leq z_{\max}, \forall n \in {\mathcal N}  \tag{\ref{SecrecyFunc}{g}}, \label{SecrecyFuncSubg}\\
    &D_g \leq D_{g,\rm max}, \forall g \in {\mathcal G} \tag{\ref{SecrecyFunc}{h}}, \label{SecrecyFuncSubh}\\
    &Q_g \geq Q_{g,\rm min}, \forall g \in {\mathcal G} \tag{\ref{SecrecyFunc}{i}}, \label{SecrecyFuncSubi}
\end{align}
where $t$ denotes time index of slot $t$, UAV-user pairing variable is $\mathcal{R} = \{r_{k, n}, \forall n \in {\mathcal N}, \forall k \in {\mathcal K}\}$, UAV trajectory planning variable is $\mathcal{O} = \{o_{n}, \forall n \in {\mathcal N}\}$, UAV-LEO pairing variable is $\mathcal{L} = \{l_{n, m}, \forall n \in {\mathcal N}, \forall m \in {\mathcal M}\}$, ISL transmission variable is $\mathcal{Y} = \{y_{i,j}, \forall i \in {\mathcal K}, \forall j \in {\mathcal K}, i \neq j\}$, transmission mode selection variable is $\varepsilon = {\varepsilon_1, \varepsilon_2, \varepsilon_{3,1}, \varepsilon_{3,2}, \varepsilon_{3,3}}$, where $\varepsilon_1$ denotes bit transmission Mode 1, $\varepsilon_2$ denotes semantic Mode 2. $\varepsilon_{3,1}$, $\varepsilon_{3,2}$ and $\varepsilon_{3,3}$ denote the three different levels of compression and image reconstruction quality in semantic Mode 3, named Mode 3\_1, Mode 3\_2 and Mode 3\_3. To be specific, in the three modes of Mode 3, the more compressed the semantic information, the lower the quality of image reconstruction. \eqref{SecrecyFuncSuba} and \eqref{SecrecyFuncSubb} indicates that a LEO can transmit signals to several UAVs, but a UAV can only connect to one LEO at a given time slot. \eqref{SecrecyFuncSubc} and \eqref{SecrecyFuncSubd} presents that a UAV can send signals to several ground users, but one user can only be served by one UAV. \eqref{SecrecyFuncSube} denotes ISLs can only be used once for a LEO at a given time slot. \eqref{SecrecyFuncSubf} shows that the UAV's speed cannot exceed the maximum limitation $v_{\max}$. The altitude constraints for UAVs are presented in \eqref{SecrecyFuncSubg}. \eqref{SecrecyFuncSubh} presents that the delay of each transmission task should be smaller than the required delay constraint $D_{g,\rm max}$, \eqref{SecrecyFuncSubi} indicates that the reconstruction quality of each task should be higher than required constraint $Q_{g,\rm min}$. Considering the time-varying Rician fading channels and dynamic locations of UAVs and LEO satellites in SAGINs, along with the uncertainty in task arrivals and requirements, this constitutes a long-term optimization problem. To achieve the optimal performance in such a dynamic environment, we utilize a DRL algorithm to learn the strategy aimed at maximizing the SCE and provide an adaptive semantic communication optimization solution for dynamic SAGINs.

\section{DRL-Based Resource Allocation} \label{sec:DSAC}
Before employing the DRL algorithm to solve the optimization problem, we first need to model the problem as a Markov Decision Process (MDP). The essential components of an MDP include the state space, action space, and reward function. We begin by defining the state at a given time slot $t$ as
\begin{equation}\small\label{eq:state}
\begin{aligned}
\jmath(t) =&~\{t, \{o_n(t)\}_{o_n \in \mathcal O}, \{h_{n, m}(t)\}_{n \in \mathcal N, m \in \mathcal M}, \\
     &~\{Q_g(t)\}_{g \in \mathcal G}, D_g(t)\}_{g \in \mathcal G}\}.
\end{aligned}
\end{equation}
Furthermore, we assume that action $a(t)$ is expressed as
\begin{equation}\small\label{eq:action}
\begin{aligned}
a(t) = \{ \mathcal{R}(t), {\mathcal O}(t), \mathcal{L}(t), \mathcal{Y}(t), \varepsilon(t) \}.
\end{aligned}
\end{equation}
Moreover, The reward function is utilized to train the agent in DRL to learn the relationships between action space and state space. We define the reward function as
\begin{equation}\small\label{eq:reward}
\begin{aligned}
r(t) = \sum_{g=1}^{G} \varpi_g(t) - \varphi,
\end{aligned}
\end{equation}
where $\varphi = 1$ if any constraint is not satisfied, otherwise $\varphi = 0$. Based on this reward function, the agent in the DRL framework will aim to maximize the cumulative SCE, which equates to maximizing the average SCE value.

To learn an effective resource allocation strategy that maximizes cumulative SCE, and to consider the relationship between cumulative SCE returns and the DRL's expectations, we use the DSAC algorithm in this work. This algorithm integrates an analysis and consideration of reward distribution into the SAC framework. The reward update process in DSAC is defined as
\begin{equation}\small\label{eq:dsac1}
\begin{aligned}
\wp^\Re(\jmath(t),a(t)) = r(t) + \hbar \vartheta(t+1),
\end{aligned}
\end{equation}
where $\Re$ indicates the agents' current policy, $\hbar$ denotes the discount in DRL trainning, $\vartheta(i) = \sum_{i=t}^{\infty} \hbar^(i-t) [ r(i) - \iota {\rm log} \Re (a(i)|\jmath(i)) ]$, where $\iota$ presents a factor in SAC \cite{9448360}. In DSAC, we can obtain the state-action value as
\begin{equation}\small\label{eq:dsac2}
\begin{aligned}
J^{\Re}(\jmath(t),a(t)) = \mathbb{E} \big[\wp^\Re(\jmath(t),a(t))\big],
\end{aligned}
\end{equation}
where ${\mathbb E}[\cdot]$ represents the expectation. Instead of focusing on the expected state-action returns as in \eqref{eq:dsac2}, we directly utilize the soft returns in \eqref{eq:dsac2} to develop the algorithm. Incorporating the principle of maximum entropy in this context, we define the distribution for the Bellman operator as
\begin{equation}\small\label{eq:dsac3}
\begin{aligned}\Im
\Im \wp^\Re(\jmath(t),a(t)) = &~r(t) + \hbar \big( \wp^\Re(\jmath(t+1),a(t+1)) \\
&- \nu {\rm log} \Re (a(t+1)|\jmath(t+1))\big).
\end{aligned}
\end{equation}
Thus, the soft return distribution can be updated according to \eqref{eq:dsac3} and can be expressed as
\begin{equation}\small\label{eq:dsac4}
\begin{aligned}
\hat{\wp}_{new} = {\rm argmin} \mathbb{E} \big[ d_s (\Im \hat{\wp}_{o}(\cdot|\jmath(t),a(t)), \hat{\wp}(\cdot|\jmath(t),a(t)) ) \big],
\end{aligned}
\end{equation}
where $d_s$ is the distance between new and old distributions, which can be expressed by Kullback–Leibler (KL) divergence \cite{pmlrv70bellemare17a}. To update the soft return strategy, we form the function as
\begin{equation}\small\label{eq:dsac5}
\begin{aligned}
\upsilon_{\hat{\wp}}(\Theta) = - \mathop{\mathbb{E}}\limits_{\Phi_{\upsilon_{\hat{\wp}}(\Theta)}} \Big[{\rm log} \aleph \big(Psi^{\Re_{\phi^{\Upsilon}}} \wp(\jmath(t),a(t)) | \hat{\wp}(\cdot|\jmath(t),a(t)) \big)   \Big],
\end{aligned}
\end{equation}
where $\Theta$ denotes a factor in updating function, $\phi$ presents policy $\Re_\phi(\cdot|s)$, $\Phi_{\upsilon_{\hat{\wp}}(\Theta)} = (\jmath(t),\jmath(t+1),a(t),r(t))\sim \mathbb{B}, a(t+1)\sim \Re_{\phi_{\Upsilon}}, \wp(\jmath(t+1),a(t+1))\sim \hat{\wp}_{\Theta^{\Upsilon}}(\cdot|\jmath(t+1),a(t+1))$, $\aleph$ represents the probability distribution over state-action pairs, while $\mathbb{B}$ refers to the sample buffer. Parameters $\Theta_{\phi^{\Upsilon}}$ and $\phi_{\phi^{\Upsilon}}$ denote the target return and target strategy, respectively. Subsequently, parameter $\Theta$ can be updated by
\begin{equation}\small\label{eq:dsac6}
\begin{aligned}
\digamma_\Theta \upsilon_{\hat{\wp}}(\Theta) = &~- \mathop{\mathbb{E}}\limits_{\Phi_{\digamma_\Theta \upsilon_{\hat{\wp}}(\Theta)}} \Big[\digamma_\Theta {\rm log} \aleph \big(Psi^{\Re_{\phi^{\Upsilon}}} \wp(\jmath(t),a(t)) \\
&~| \hat{\wp}(\cdot|\jmath(t),a(t)) \big)   \Big],
\end{aligned}
\end{equation}
where $\Phi_{\upsilon_{\hat{\wp}}(\Theta)} = (\jmath(t),\jmath(t+1),a(t),r(t))\sim \mathbb{B}, a(t+1)\sim \Re_{\phi_{\Upsilon}}, \wp(\jmath(t+1),a(t+1))\sim \hat{\wp}_{\Theta^{\Upsilon}}$. $\hat{\wp}_{\Theta}$ has random distribution. Then, the updating function is given by
\begin{equation}\small\label{eq:dsac7}
\begin{aligned}
\digamma_\Theta \upsilon_{\hat{\wp}}(\Theta) = &~\mathop{\mathbb{E}}\limits_{\Phi_{\digamma_\Theta \upsilon_{\hat{\wp}}(\Theta)}} \Big[ - \frac{\nabla \Psi_{\hat{\wp}}(\Theta)}{\nabla J_{\Theta}(\jmath(t),a(t))} \digamma_{\Theta}J_{\Theta}(\jmath(t),a(t)) \\
&~- \frac{\nabla \Psi_{\hat{\wp}}(\Theta)}{\nabla \prime_{\Theta}(\jmath(t),a(t))} \digamma_{\Theta}\prime_{\Theta}(\jmath(t),a(t)) \Big],
\end{aligned}
\end{equation}
where
\begin{equation}\small\label{eq:dsac8}
\begin{aligned}
\frac{\nabla \Psi_{\hat{\wp}}(\Theta)}{\nabla J_{\Theta}(\jmath(t),a(t))} = \frac{\Im^{\Re_{\phi_{\Upsilon}}}\wp(\jmath(t),a(t)) - J_{\Theta}(\jmath(t),a(t)) }{\prime_{\Theta}(\jmath(t),a(t))^2},
\end{aligned}
\end{equation}
\begin{equation}\small\label{eq:dsac9}
\begin{aligned}
\frac{\nabla \Psi_{\hat{\wp}}(\Theta)}{\nabla \prime_{\Theta}(\jmath(t),a(t))} = &~\frac{\Im^{\Re_{\phi_{\Upsilon}}}\wp(\jmath(t),a(t)) - J_{\Theta}(\jmath(t),a(t)) }{\prime_{\Theta}(\jmath(t),a(t))^3} \\
&~- \frac{1}{\prime_{\Theta}(\jmath(t),a(t))},
\end{aligned}
\end{equation}
where $\prime_{\Theta}(\jmath(t),a(t))$ denotes the standard deviation of the distribution for $\wp(\cdot|\jmath(t),a(t))$. According to \eqref{eq:dsac9}, we can see that the value is prone to overestimation during training process. Besides, when the standard deviation approaches either 0 or infinity, challenges can emerge in computing gradients. Thus, we constrain the standard deviation by
\begin{equation}\small\label{eq:dsac10}
\begin{aligned}
\prime_{\Theta}(\jmath(t),a(t)) = \max \Big( \prime_{\Theta}(\jmath(t),a(t)), \prime_{\min}  \Big),
\end{aligned}
\end{equation}
where $\prime_{\min}$ denotes the parameter of constrained updating. We can also utilize the clip algorithm to constrain the updating by using equation as
\begin{equation}\small\label{eq:dsac11}
\begin{aligned}
\frac{\nabla \Psi_{\hat{\wp}}(\Theta)}{\nabla \prime_{\Theta}(\jmath(t),a(t))} = &~\frac{s - J_{\Theta}(\jmath(t),a(t)) }{\prime_{\Theta}(\jmath(t),a(t))^3} \\
&~- \frac{1}{\prime_{\Theta}(\jmath(t),a(t))},
\end{aligned}
\end{equation}
where
\begin{equation}\small\label{eq:dsac12}
\begin{aligned}
s = \ell \Big(\Im^{\Re_{\phi_{\Upsilon}}}\wp(\jmath(t),a(t)), J_{\Theta}(\jmath(t),a(t)) - \partial, J_{\Theta}(\jmath(t),a(t)) + \partial \Big),
\end{aligned}
\end{equation}
where $\ell[l,i,j]$ presents that $l$ is clipped between $[i, j]$, $\partial$ denotes the parameter for updating. Thus, we can update the DSAC agent's strategy as
\begin{equation}\small\label{eq:dsac13}
\begin{aligned}
\upsilon_{\Re}(\phi) = \mathop{\mathbb{E}}\limits_{\jmath(t)\sim \mathbb{B},a(t)\sim \Re_\phi} \Big[ J_{\Theta}(\jmath(t),a(t)) - \nu {\rm log}(\Re_{\phi}(a(t)|\jmath(t))) \Big].
\end{aligned}
\end{equation}
Policy $\phi$ can be updated based on
\begin{equation}\small\label{eq:dsac14}
\begin{aligned}
\digamma_\phi \upsilon_{\Re}(\phi) = &~\mathop{\mathbb{E}}\limits_{\jmath(t)\sim \mathbb{B},\rho} \Big[- \nu \digamma_\phi{\rm log}(\Re_{\phi}(a(t)|\jmath(t))) \big(\digamma_{a(t)}J_{\Theta}(\jmath(t),\\
&~ a(t))- \nu \digamma_{a(t)}{\rm log}(\Re_\phi(a(t)|\jmath(t)))\big) \digamma_\phi \imath(\rho;\jmath(t)) \Big],
\end{aligned}
\end{equation}
where $\imath(\rho;\jmath(t)) = \hat{a} + \rho \bigodot \bar{a}$, $\rho$ represents a sample obtained from a fixed set of random sampling, $\bigodot$ indicates the Hadamard product, $\hat{a}$ and $\bar{a}$ present the mean and standard deviation of $\Re_{\phi}(\cdot|\jmath(t))$, respectively.

\section{Simulation Results} \label{sec:sim}
\begin{figure*}[t!]
    \centering
    \includegraphics[width=0.95\linewidth]{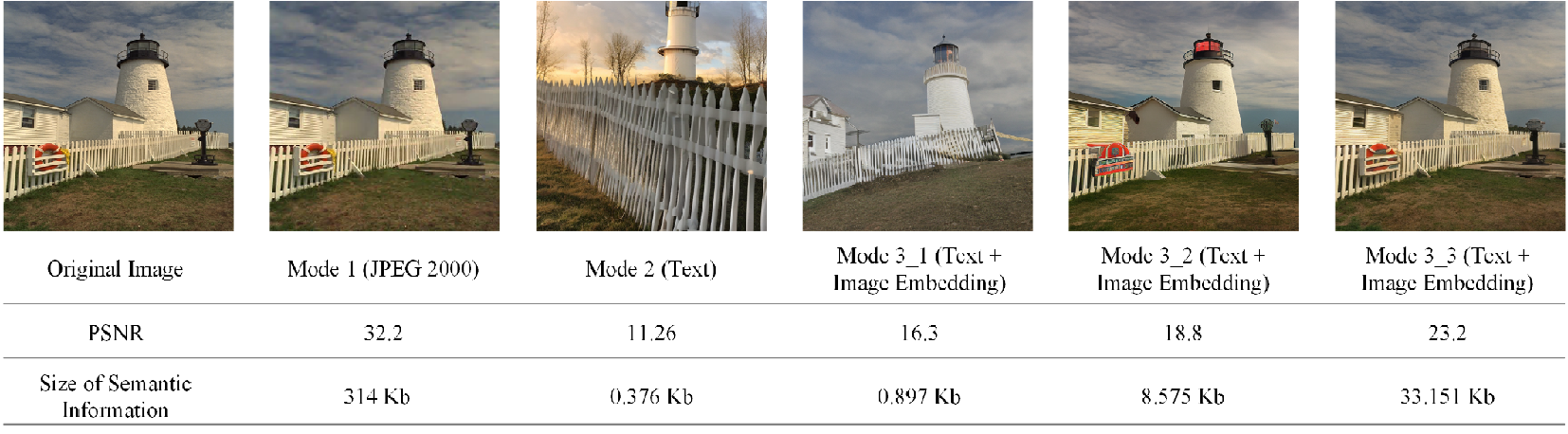}
    \caption{The reconstruction quality in our hybrid bit and semantic communication framework.}
    \label{fig:Rs}
\end{figure*}
Unless otherwise stated, simulation parameters are set as follows: the number of ground users $M = 8$, the number of UAVs $N = 3$, the number of LEO satellites $K = 5$, ground users are randomly distributed in $N$ areas, a UAV is randomly deployed in each area and the initial altitude is 50 m for all UAVs. The altitude of LEOs is 750 km, the distance between LEO satellites are randomly generated from 250 km to 500 km. The speed of LEOs is 7.8 km/s, the maximum speed of UAVs is 10 m/s, the maximum and minimum UAV altitudes are 60m and 45m, respectively. The antenna gain of LEOs $\varsigma = 40$ dB, the maximum Doppler frequency is based on Ka-band as in \cite{10043628}, the carrier frequency of LEO transmission is 25 GHz, the transmit powers of LEOs and UAVs are 1 W and 0.2 W, respectively. Thermal noise $\chi = 354.81$ K \cite{ISL_TWC21}, the noise level at UAVs and ground users is -130 dBm, the minimum elevation angle $\tau = 40 ^{\circ}$, the bandwidth for all transmissions is 10 MHz, the path loss exponents $\kappa_L = 2$ and $\kappa_N = 2.6$. $\hbar = 0.99$, $\prime_{\min} = 1$, image transmission tasks arrive following FTP Model 3 \cite{3GPPFTPmodel3} at each LEO. The size of an original image is 3.5 MB, the number of total tasks is $G = 100$. The weights of SCE are 0.5. The delay constraint for each task is randomly distributed between 2 s and 11 s, the quality constraint PSNR threshold for each task is randomly distributed between 10 and 30. In our hybrid bit and semantic communication framework, we utilize JPEG 2000 as the traditional image compression algorithm in bit transmission Mode 1. The computation delay for Mode 1 is 0.1323 s, for other modes is 1.33 s. Besides, the multi-agent proximal policy optimization (MAPPO) algorithm is adopted as the benchmark in simulation.

From Fig. \ref{fig:Rs}, the traditional compression method (JPEG 2000) via bit transmission mode (Mode 1) yields the image reconstruction quality closest to the original at the ground user. Furthermore, the traditional compression approach requires minimal computation time, thereby having very little impact on computational delay. However, the main disadvantage of the traditional compression algorithm is its substantial transmission resource requirements, which leads to challenges for satellite link budgets. In contrast, the multiple adaptive semantic transmission modes of our proposed transmission framework can flexibly achieve a favorable balance between performance and transmission needs. Nonetheless, computational delay remains a significant factor that cannot be overlooked in semantic communication.

\begin{figure}[t!]
  \centering
  \centerline{\includegraphics[scale=0.55]{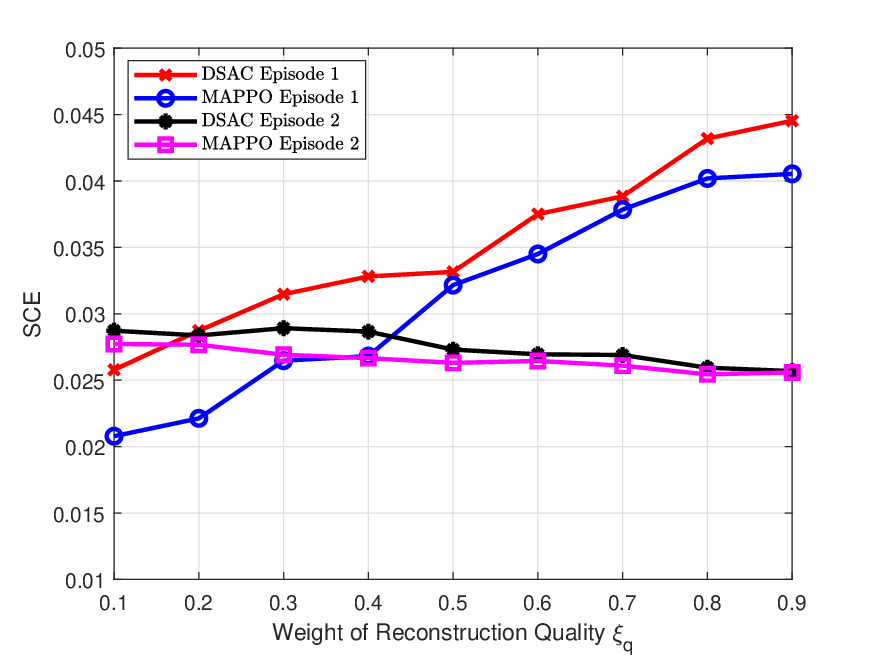}}
 \caption{\small Average SCE vs the weight of reconstruction quality at ground user.} \label{fig:R1}
\end{figure}

In Fig. \ref{fig:R1}, we compare the SCE performance difference between two episodes (i.e., two runs after random re-initialization). From the figure, the proposed DSAC-based method performs better than the benchmark. In Episode 1, the average SCE value increases as the reconstruction quality weight increases, because there are many selections of high-quality transmission modes in this particular episode, such as the bit transmission mode (Mode 1) and semantic Mode 3\_3. As a result, the rise in reconstruction quality weight had a positive effect on SCE. In Episode 2, the SCE showed a slight decrease as the reconstruction weight increased since there are more selections of modes with lower reconstruction quality but lower transmission delay, such as Mode 2. These simulation results demonstrate that the proposed metric can be effectively used to analyze the trade-off between transmission and delay in the SAGIN scenario, and can further support the optimization of using adaptive hybrid bit and semantic transmission.

\begin{figure}[t!]
  \centering
  \centerline{\includegraphics[scale=0.55]{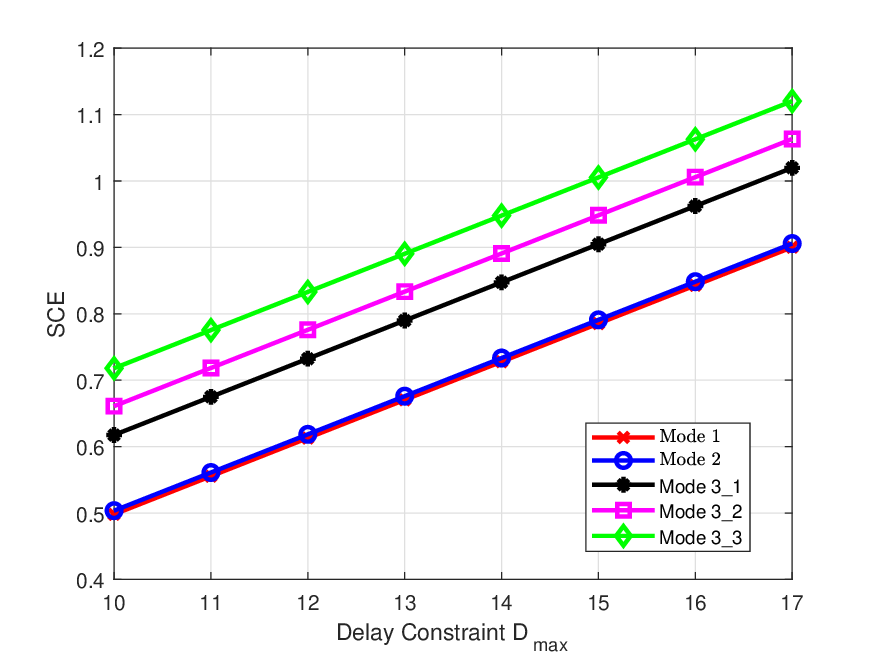}}
 \caption{\small Average SCE vs delay constraint $D_{\rm max}$.} \label{fig:R2}
\end{figure}

To analyze and compare the proposed transmission modes, we conduct an experiment, where we set the delay and reconstruction quality constraints of all tasks to be consistent and adjust their values so all tasks can be completed by using any transmission mode. The reconstruction quality constraint is set to 10, slightly below the lowest PSNR value of any transmission mode. We compare how changes in the delay constraint affected the SCE. As shown in Fig. \ref{fig:R2}, with an increase in the delay constraint, the average SCE of all transmission modes also increased, because a higher delay constraint leads to a significant SCE gain as defined in \eqref{metric}. Furthermore, Mode 3\_3 achieves the highest SCE value in this episode because it can balance the transmission time and reconstruction quality. Since bit transmission requires the most link resource although has significant advantages in terms of computation delay, it has substantial disadvantages in terms of transmission time, resulting in worse performance when using pure bit transmission. Notice that the SCE values in the results are significantly higher compared to those in Fig. \ref{fig:R1}, because we significantly increase the delay constraints for all tasks to ensure that each mode could individually complete the task. However, this increase in delay constraints also substantially enhance the SCE values based on \eqref{metric}.

\begin{figure}[t!]
  \centering
  \centerline{\includegraphics[scale=0.55]{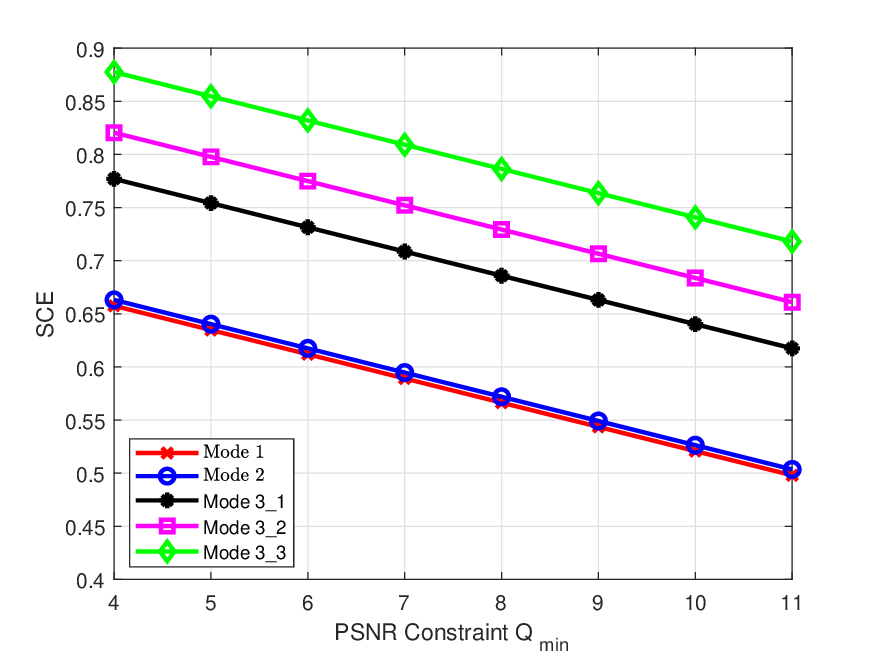}}
 \caption{\small Average SCE vs reconstruction quality $Q_{\rm min}$.} \label{fig:R3}
\end{figure}

Similar to the experiment shown in Fig. \ref{fig:R2}, we set the delay constraint to 11 sec and compare the impact of different reconstruction quality constraints on the final performance. From Fig. \ref{fig:R3}, all values follow the setup in \eqref{metric}, where the SCE decreases as the performance constraint increases. Furthermore, Mode 3\_3 once again proves to be the most balanced option.

\begin{figure}[t!]
  \centering
  \centerline{\includegraphics[scale=0.55]{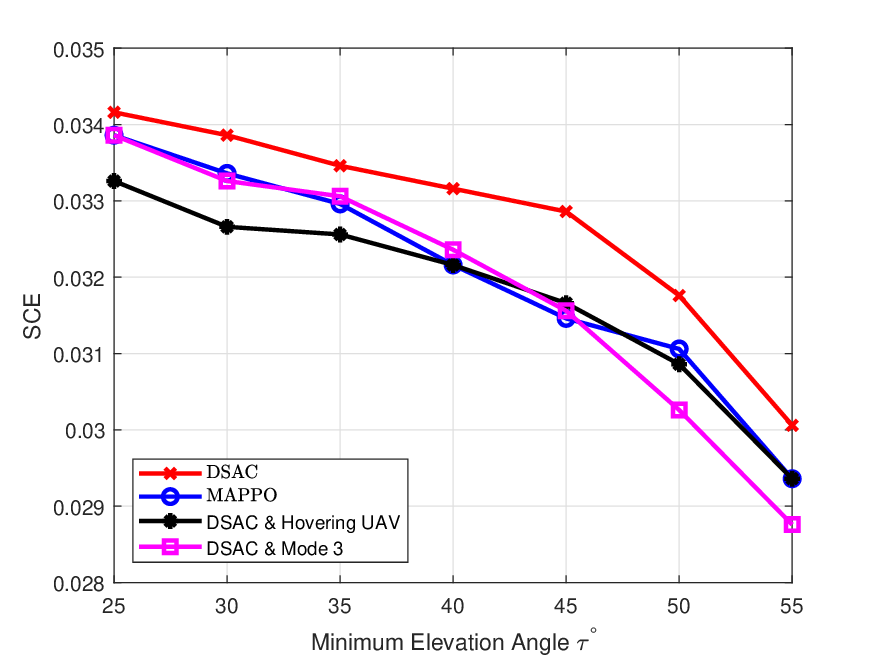}}
 \caption{\small Average SCE vs minimum elevation angle.} \label{fig:R4}
\end{figure}

In Fig. \ref{fig:R4}, we analyze the performance of the proposed algorithm compared with other benchmark schemes under different minimum elevation angles. As shown in this figure, the minimum elevation angle impacts the connection time between the LEO satellites and UAVs. Therefore, as the minimum elevation angle increases, the LEO satellite is forced to pass on part of the tasks to other satellites, introducing additional delays. `DSAC \& Hovering UAV' represents a scheme without UAV trajectory planning, whereas `DSAC \& Mode 3' represents a method without semantic Mode 3\_1 and Mode 3\_2 options. The results illustrate that UAV trajectory planning provides significant gains to the proposed system while the inclusion of more semantic transmission modes offers greater flexibility in balancing transmission delay and performance for the image transmission tasks.

\begin{figure}[t!]
  \centering
  \centerline{\includegraphics[scale=0.55]{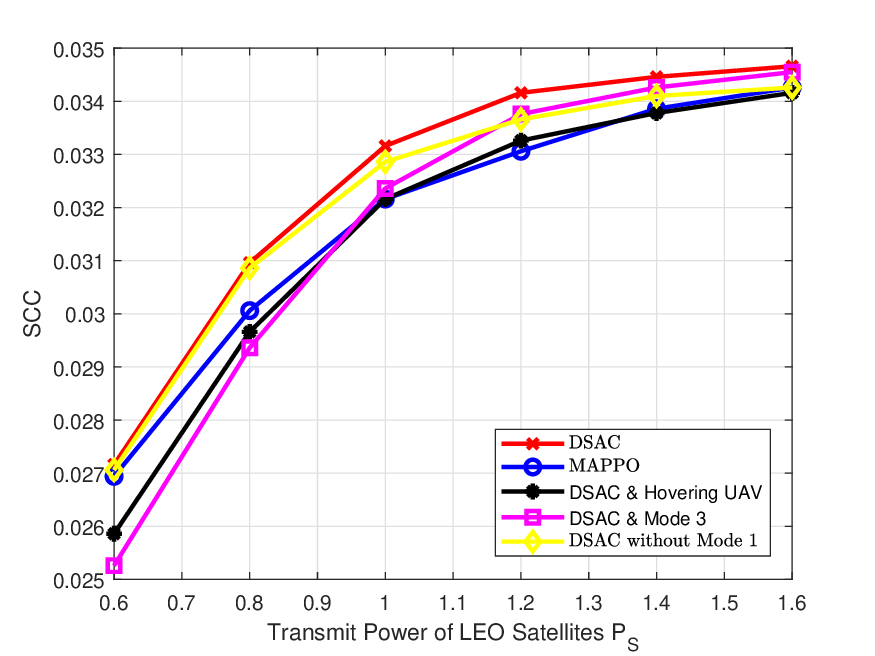}}
 \caption{\small Average SCE vs the transmit power of LEO satellites $P_S$.} \label{fig:R5}
\end{figure}

Considering that the link budget in the proposed system is closely related to the satellite transmit power, we demonstrate the relationship between satellite transmit power and SCE in Fig. \ref{fig:R5}. From the figure, higher satellite transmit power can support greater performance gain because increased power provides sufficient link budget, and allows the transmission mode to prefer selecting high-quality generative modes while maintaining delay metrics, or to complete transmissions faster under the same reconstruction quality, thereby reducing delays and improving the overall SCE. Moreover, `DSAC without Mode 1' represents a purely semantic transmission framework excluding Mode 1. When satellite transmit power is low, the absence of bit transmission has little impact, as the low SNR communication condition is insufficient to support the high data rate required for bit transmissions. However, as the satellite transmit power increases, bit transmission begins to play a significant role, thereby providing greater advantages in the proposed hybrid bit and semantic transmission framework.

\begin{figure}[t!]
  \centering
  \centerline{\includegraphics[scale=0.55]{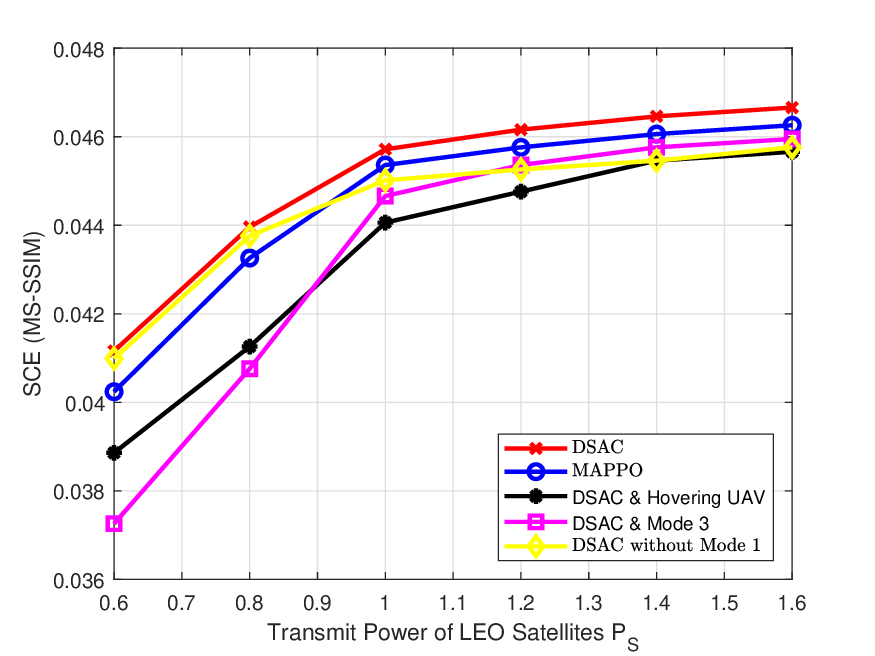}}
 \caption{\small Average SCE (MS-SSIM) vs the transmit power of LEO satellites $P_S$.} \label{fig:R6}
\end{figure}

To further analyze the robustness of the proposed semantic performance metric SCE using an alternative evaluation source, we replace the measuring method for quality of the reconstructed image at the ground user with multi-scale structural similarity index measure (MS-SSIM) in \eqref{metric}. As shown in Fig. \ref{fig:R6}, the SCE performance still increases with higher satellite transmission power and the analysis related to bit transmission is consistent with Fig. \ref{fig:R5}. Therefore, the proposed hybrid bit and semantic transmission framework demonstrates significant potential within SAGIN.

\section{Conclusion}\label{sec:con}
This paper introduced hybrid bit and semantic transmission framework within SAGINs, including UAV planning, user-UAV pairing, UAV-LEO pairing, and ISLs. In this framework, bit transmission employs the JPEG 2000 compression scheme while semantic transmission modes are categorized into two types: text transmissions and text \& image embeddings transmissions. Furthermore, the text \& image embeddings mode is further divided into three sub-modes based on transmission requirements and semantic performance, providing greater flexibility for wireless transmission choices. Moreover, our novel semantic communication metric SCE balances the semantic information quality and transmission delay. By employing DSAC algorithm, we optimized resource allocation in the SAGIN to maximize the average SCE. Simulation results validated the advantages of the proposed transmission framework and optimization algorithm, providing more flexible transmission modes and trade-off strategies for future large-scale and complex wireless communication systems. In our future work, we will investigate semantic communication frameworks under conditions where encoder and decoder have different knowledge bases, and design a hierarchical codebook structure where base and enhancement codebooks can be dynamically combined to support multiple bitrate levels.

\bibliographystyle{IEEEtran}
\bibliography{IEEEabrv,ref}
\end{document}